\documentclass[twocolumn,showpacs,preprintnumbers,amsmath,amssymb]{revtex4}
\usepackage{tabularx,graphicx}\begin{document}
%\documentstyle[aps]{revtex}
%\documentstyle[preprint,aps]{revtex}
%\begin{document}
\newcommand{\beq}{\begin{equation}}
\newcommand{\eeq}{\end{equation}}
\newcommand{\beqn}{\begin{eqnarray}}
\newcommand{\eeqn}{\end{eqnarray}}
\newcommand{\bmath}{\begin{subequations}}
\newcommand{\emath}{\end{subequations}}
\newcommand{\bk}{\bold{k}}
\newcommand{\bkp}{\bold{k'}}
\newcommand{\bq}{\bold{q}}
\newcommand{\br}{\bold{r}}
\newcommand{\up}{\uparrow}
\newcommand{\dn}{\downarrow}
\newcommand{\ct}{\tilde{c}}
\newcommand{\vt}{\tilde{V}}
\newcommand{\tk}{\tilde{\bk}}

%\draft
\title{A new basis set for the description of electrons in  superconductors}
\author{J. E. Hirsch }
\address{Department of Physics, University of California, San Diego\\
La Jolla, CA 92093-0319}
 
%\date{April 13, 2009} 
\begin{abstract} 

In the usual description of electrons in metals and superconductors, the single electron states are assumed to satisfy Bloch's theorem. This is because the electron-ion interaction is privileged over the electron-electron interaction. However   the theory of hole superconductivity proposes that in the transition to superconductivity  carriers `undress' from the electron-ion interaction.
I propose here a new basis set to describe electrons in the superconducting state 
that does not satisfy Bloch's theorem but instead is designed to optimize the electron-electron interaction.
The new basis set favors a superconducting state where a spin current exists and where states near the bottom of the band become partially occupied, as predicted  by the theory of hole superconductivity.

 \end{abstract}
\pacs{}
\maketitle

\section{Introduction}

It is generally assumed that electron states in a periodic solid obey Bloch's theorem\cite{bloch}. That is, a wavevector $\bold{k}$  characterizes the single electron orbital $\psi_{n\bold{k}} (\bold{r} )$ which obeys the relation

\beq
\psi_{n\bold{k}}(\bold{r}+\bold{R})=e^{i\bold{k}\cdot \bold{R}} \psi_{n\bold{k}}(\bold{r})
\eeq
where the index $n$ (band index) differentiates different states with the same Bloch wavevector $\bold{k}$. In Eq. (1), $\bold{R}$ is any lattice translation vector
that leaves the electron-ion potential invariant: 
\beq
U_{e-i}(\bold{r}+\bold{R})=U_{e-i}(\bold{r})   .
\eeq
Electron-electron interactions are assumed to convert the single-particle electron-ion potential to a new `effective' potential $U_{eff}$  that includes electron-electron interaction direct, exchange and correlation effects (e.g. the Kohn-Sham potential in density-functional theory)\cite{kohnsham}. The
effective potential still obeys
\beq
U_{eff}(\bold{r}+\bold{R})=U_{eff}(\bold{r})
\eeq
with the same lattice vectors $\bold{R}$, and the corresponding single particle orbitals (e.g. Kohn-Sham orbitals) still obey Bloch's theorem Eq. (1).

Similarly, in the generally accepted conventional theory of superconductivity (BCS theory)\cite{bcs} the many-body wavefunction is written as
\beq
|\Psi>_{BCS}=\prod_{\bold{k}} (u_{\bold{k}}+v_{\bold{k}} c^\dagger_{\bold{k}\uparrow} c^\dagger_{\bold{-k}\downarrow}) |0>
\eeq
where the index $\bold{k}$ is the Bloch wavevector of the state where the single electron  of spin $\sigma$ is  created by the operator  $c^\dagger_{\bold{k}\sigma}$. The normal state of the system can be expressed
in the form Eq. (4) with $u_{\bold{k}} v_{\bold{k}}=0$ and $v_{\bold{k}}=0$ or $u_{\bold{k}}=0$ depending on whether $\epsilon_{\bold{k}}<\epsilon_F$ or $\epsilon_{\bold{k}}>\epsilon_F$ with $\epsilon_F$ the Fermi energy.
The superconducting state instead is characterized by $u_{\bold{k}} v_{\bold{k}}\neq 0$ for some values of $\bold{k}$. For the case of nearly free electron metals, the Bloch states are often simply taken to be free electron plane wave states.

However, it is not necessarily true that electron states in a periodic electron-ion potential   have to obey Bloch's theorem. Consider a case where the electron-electron interaction completely overwhelms
the electron-ion interaction. The electronic spatial arrangement that minimizes the energy may be an electronic lattice (Wigner crystal, incommensurate charge density wave) with a lattice structure that has no relation with the underlying
ionic structure. For example, the electronic structure may be hexagonal and the underlying ionic lattice cubic. In such situation the single particle states describing the electrons will $not$ obey Bloch's theorem
Eq. (1).

We have recently argued  that  experimental evidence clearly indicates that in the transition to superconductivity the  carriers undergo a wavelength expansion and end up completely  `undressed' from the electron-ion interaction, 
so that they  no longer `see' the discrete nature of the electron-ion potential\cite{holeelectron2,sm}.
It is natural to conclude
that the electron states describing the superfluid will no longer obey Bloch's theorem. 

Furthermore, we have recently proposed that the superconducting state   should be understood as a (distorted) `mirror image' of the normal state, where the `mirror' switches the role
of the electron-ion and electron-electron interactions\cite{holeelectron3}. We argued that in  the superconducting state, the electron-electron interaction is privileged over the electron-ion interaction and
the many-body wavefunction optimizes the electron-electron interaction, with the electron-ion interaction being non-optimized. It is natural that such a state should be 
described by a single electron basis that does not satisfy Bloch's theorem. But what is the optimal basis to describe such  state?

The new single-particle basis states that we propose here  are defined by the relation
\bmath
\beq
\tilde{\varphi}_{\bold{k}}(\bold{r})=\frac{\varphi_\bold{k}(\bold{r}) + i\varphi_\bold{-k}(\bold{r})}{\sqrt{2}}
\eeq
where $\varphi_\bold{k}$ are the usual Bloch states for a given band. Note that $\{\tilde{\varphi}_\bold{k}\}$ is a complete orthonormal set if and only if $\{\varphi_\bold{k}\}$ is. The inverse relation is
\beq
\varphi_{\bold{k}}(\bold{r})=\frac{\tilde{\varphi}_\bold{k}(\bold{r}) - i\tilde{\varphi}_\bold{-k}(\bold{r})}{\sqrt{2}}
\eeq
\emath
Similarly we define the electron creation operators in the new states $\tilde{c}^\dagger_{\bold{k}\sigma}$ in terms of the Bloch creation operators $c^\dagger_{\bold{k}\sigma}$ by
\bmath
\beq
\tilde{c}^\dagger_{\bold{k}\sigma}=\frac{c^\dagger_{\bold{k}\sigma}+ic^\dagger_{\bold{-k}\sigma}}{\sqrt{2}}
\eeq
 \beq
c^\dagger_{\bold{k}\sigma}=\frac{\tilde{c}^\dagger_{\bold{k}\sigma}-i\tilde{c}^\dagger_{\bold{-k}\sigma}}{\sqrt{2}}
\eeq
\emath
It is clear that our new basis functions do not satisfy Bloch's theorem, since
\beqn
\tilde{\varphi}_{\bold{k}}(\bold{r}+\bold{R})&=&\frac{e^{i\bold{k}\cdot \bold{R}}\varphi_\bold{k}(\bold{r}) + ie^{-i\bold{k}\cdot \bold{R}}\varphi_\bold{-k}(\bold{r})}{\sqrt{2}} \nonumber  \\
&\neq& e^{i\bold{k}\cdot \bold{R}}\tilde{\varphi}_{\bold{k}}(\bold{r})
\eeqn
Nevertheless, I will argue that the new basis has very interesting properties and is relevant to the description of the superconducting state.

\section{Nature of the wave functions}
Consider the case where the Bloch functions are well approximated by plane wave states
\beq
\varphi_\bold{k}(\bold{r})=\frac{1}{\sqrt{\Omega}} e^{i\bold{k}\cdot\bold{r}}
\eeq
Then
\bmath
\beq
\tilde{\varphi}_\bold{k}(\bold{r})=\frac{1+i}{\sqrt{\Omega}}cos(\bold{k}\cdot\bold{r}-\frac{\pi}{4})
\eeq
\beq
\tilde{\varphi}_{-\bold{k}}(\bold{r})=\frac{1+i}{\sqrt{\Omega}}cos(\bold{k}\cdot\bold{r}+\frac{\pi}{4}) .
\eeq
\emath
describing standing waves. The values of $\bold{r}$ for which $|\tilde{\varphi}_\bold{k}(\bold{r})|^2$ is maximum satisfy
\bmath
\beq
\bold{k}\cdot\bold{r}-\frac{\pi}{4}=n\pi
\eeq
with $n$ integer, and  for those $\bold{r}$ values
\beq
\bold{k}\cdot\bold{r}+\frac{\pi}{4}=(n+\frac{1}{2})\pi
\eeq
\emath
so that   $|\tilde{\varphi}_{-\bold{k}}(\bold{r})|^2=0$. Consequently this will cause a strong reduction of the direct Coulomb repulsion between an electron in state $\tilde{\varphi}_\bk$  and another electron
in state $\tilde{\varphi}_{-\bk}$ compared to the plane wave case where the density  is uniform,
as illustrated in Fig. 1. Conversely, the in-phase modulation of the wavefunction amplitude for two electrons in the $same$ state
$\tilde{\varphi}_\bold{k}$ will give a strong enhancement of the  direct Coulomb repulsion in that case  compared to the plane wave case. 

 \begin{figure}
 \resizebox{6.5cm}{!}{\includegraphics[width=7cm]{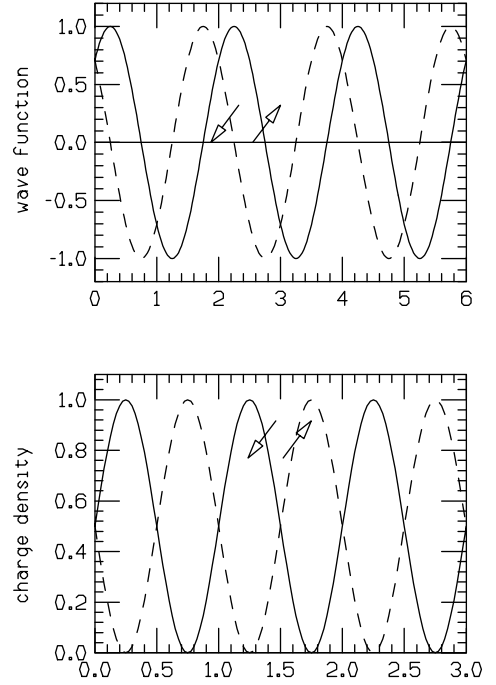}}
 \caption {Wave function amplitudes and charge densities for the standing waves associated with the operators  $\tilde{c}_{\bk \sigma}^\dagger$ (full lines) and
 $\tilde{c}_{-\bk,  -\sigma}^\dagger$ (dashed lines) , assuming the Bloch states are plane waves. Note that the charge densities avoid each other reducing the Coulomb repulsion.
}
 \label{figure2}
 \end{figure}

Similarly, consider a band described by the tight binding approximation with a single orbital $\phi_i(\bold{r})$ per site. The Bloch functions are
\beq
\varphi_\bk (\br)=\frac{1}{\sqrt{N}} \sum_i e^{i\bk\cdot\bold{R}_i}\phi_i(\br)
\eeq
and the new orbitals are given by
\bmath
\beq
\tilde{\varphi}_\bk(\br)=\frac{1+i}{\sqrt{N}} \sum_i cos(\bk\cdot\bold{R}_i-\frac{\pi}{4})\phi_i(\br)
\eeq
\beq
\tilde{\varphi}_{-\bk}(\br)=\frac{1+i}{\sqrt{N}} \sum_i cos(\bk\cdot\bold{R}_i+\frac{\pi}{4})\phi_i(\br)
\eeq
\emath
It is clear that for small $\bk$ the situation is the same as for the plane-wave case: the values of $\bold{R}_i$ for which the amplitude is maximum in Eq. (12a) coincide (or nearly coincide) with 
those where the amplitude in Eq. (12b) is zero. However, this changes when the wavevector $\bk$ becomes an appreciable fraction of a reciprocal lattice vector $\bold{G}$. We have
\beq
 cos(\bk\cdot\bold{R}_i\mp\frac{\pi}{4})=\frac{1}{\sqrt{2}}  (cos(\bk\cdot\bold{R}_i)\pm sin(\bk\cdot \bold{R}_i))
\eeq 
so that the amplitudes in Eq. (12) become the same when $\bk\rightarrow \bold{G}/2$, since the second term in Eq. (13) vanishes.
Thus, we expect the direct Coulomb repulsion between an electron in state $\tilde{\varphi}_\bk$ and one in state $\tilde{\varphi}_{-\bk}$ to be reduced, and that between two electrons in the same state
$\tilde{\varphi}_\bold{k}$ to be enhanced, when $\bk$ is near the bottom of the band but not when $\bk$ is near the top of the band.

\section{Coulomb repulsion in the new basis}

The bare Coulomb interaction between opposite spin electrons in a plane wave basis is given by
\bmath
\beq
V_c=\frac{1}{\Omega} \sum_{\bk \bk' \bq} V(\bq)c^\dagger_{\bk+\bq\uparrow}c^\dagger_{\bk'-\bq\downarrow}c_{\bk' \downarrow}c_{\bk \uparrow}
\eeq
\beq
V(\bq)=\frac{4\pi e^2}{q^2}
\eeq
\emath
with $\Omega$ the volume of the system. The Coulomb  interaction between an electron of spin $\uparrow$ in the plane wave state $\varphi_\bk$ and an electron of spin $\downarrow$ in   $\varphi_\bk '$ 
\bmath 
\beq
|\bk \up ,\bk '\dn > \equiv c^\dagger_{\bk\uparrow} c^\dagger_{\bk '\downarrow} |0>
\eeq
is simply
\beqn
&<&\bk\up ,\bk ' \dn| V_c |\bk \up,\bk ' \dn>= \nonumber \\
&\int& d^3r d^3 r' |\varphi_\bk (\br)|^2  \frac{e^2}{|\br-\br'|}  |\varphi_{\bk'}(\br)|^2=       \frac{1}{\Omega} V(0)
\eeqn
\emath
independent of  $\bk$ and $\bkp$.  The Coulomb interaction between an electron of spin $\uparrow$ in state $\tilde{\varphi}_\bk$ and an electron of spin $\downarrow$ in state $\tilde{\varphi}_\bk '$:
\bmath
\beq
|\tilde{\bk}\up ,\tilde{\bk} '\dn > \equiv  \tilde{c}^\dagger_{\bk\uparrow} \tilde{c}^\dagger_{\bk '\downarrow} |0>
\eeq
is given by
\beqn
&<&\tilde{\bk}\up ,\tilde{\bk} '\dn | V_c|\tilde{\bk}\up ,\tilde{\bk} '\dn >= \frac{1}{\Omega} V(0) \nonumber \\
 &+& \frac{1}{2\Omega}V(2k)(-\delta_{\bk',-\bk}   
+  \delta_{\bk',\bk})
\eeqn
\emath
showing  indeed a reduction for $(\bk \up, -\bk \dn)$ in the new basis as well as an increased repulsion for $(\bk\up, \bk \dn)$, since the 
Coulomb matrix element $V(2k)$ is always positive. Note also that {\it the effect is largest for small k}.

Similarly consider the Coulomb interactions in the tight binding model. We will assume for simplicity that the only non-zero matrix element of the Coulomb interaction in the
tight binding model is the onsite Coulomb repulsion U:
\beq
U=\int d^3r d^3 r' |\varphi_i (\br)|^2  \frac{e^2}{|\br-\br'|}  |\varphi_i(\br ')|^2
\eeq
The Coulomb repulsion between an electron in Bloch state $\bk$ and one in Bloch state $\bk'$ is
\beq
<\bk\up ,\bk ' \dn| V_c |\bk \up,\bk ' \dn>=\frac{U}{N}
\eeq
independent of $\bk$ and $\bk'$. Instead, in the new basis
\beqn
&<&\tk\up ,\tk ' \dn| V_c |\tk \up,\tk ' \dn>= \nonumber \\
&& \frac{U}{N^2}\sum_i (1+sin(2\bk\cdot\bold{R_i})) (1+sin(2\bk'\cdot\bold{R_i}))
\eeqn
showing in particular that the interaction is most reduced for $\bk'=-\bk$, and most enhanced for $\bk'=\bk$. However, for  $\bk$ and $\bk'$ denoting states near the top of the
band, $\sim \bold{G}/2$, with $\bold{G}$ a reciprocal lattice vector,  the effect disappears, in agreement with the discussion in the previous section.

However, Eq. (19) appears to suggest that the Coulomb interactions are the same also for states around the half-filled band, e.g. $\bk=\bold{G}/4$. This is  an artifact caused by keeping only
the on-site Coulomb matrix element $U$. When considering other Coulomb matrix elements also it is found that the  Coulomb interaction between states in the new basis and in the
conventional  basis become the same only when the states approach the top of the band and the wavevectors are connected by reciprocal lattice vectors.

Next we consider the Coulomb interaction in the triplet and singlet $(\bk\uparrow,-\bk\downarrow)$ pairs. We have
\beq
\tilde{c}^\dagger_{\bold{k}\uparrow}\tilde{c}^\dagger_{-\bold{k}\downarrow}-\tilde{c}^\dagger_{-\bold{k}\uparrow}\tilde{c}^\dagger_{\bold{k}\downarrow}=c^\dagger_{\bold{k}\uparrow}c^\dagger_{-\bold{k}\downarrow}
-c^\dagger_{-\bold{k}\uparrow}c^\dagger_{\bold{k}\downarrow}
\eeq
so that the triplet state in the new basis, 
\beq
|\tilde{t}>=\frac{\tilde{c}^\dagger_{\bold{k}\uparrow}\tilde{c}^\dagger_{-\bold{k}\downarrow}-\tilde{c}^\dagger_{-\bold{k}\uparrow}\tilde{c}^\dagger_{\bold{k}\downarrow}}{2}|0>
\eeq
is identical to the triplet state in the conventional basis, $|t>$. The Coulomb interaction is
\beq
<\tilde{t}|V_c|\tilde{t}>=<t|V_c|t>=\frac{1}{\Omega}(V(0)-V(2k)) .
\eeq
The singlet states are different however, since
\beq
\tilde{c}^\dagger_{\bold{k}\uparrow}\tilde{c}^\dagger_{-\bold{k}\downarrow}+\tilde{c}^\dagger_{-\bold{k}\uparrow}\tilde{c}^\dagger_{\bold{k}\downarrow}=
i(c^\dagger_{\bold{k}\uparrow}c^\dagger_{\bold{k}\downarrow}
+c^\dagger_{-\bold{k}\uparrow}c^\dagger_{-\bold{k}\downarrow})
\eeq
and we have
\bmath
\beq
<\tilde{s}|V_c|\tilde{s}>=\frac{V(0)}{\Omega} 
\eeq
\beq
<s|V_c|s>=\frac{1}{\Omega}(V(0)+V(2k))
\eeq
\emath
so the Coulomb repulsion is smaller in the singlet state of the new basis than in the singlet state of the conventional basis.
The average of Eqs. (22) and (24a) or (24b) yields  Eq. (16b) or Eq. (15b) for $\bk'=-\bk$, since the state $|\bk \uparrow, -\bk\downarrow>$ is a linear combination of singlet and triplet states with equal weight.

A similar result holds in the tight binding model, where we have (assuming only on-site interactions)

\bmath
\beq
<\tilde{t}|V_c|\tilde{t}>=<t|V_c|t>=0
\eeq
\beq
<\tilde{s}|V_c|\tilde{s}>=\frac{2U}{N^2} \sum_i cos^2 2\bk\cdot\bold{R}_i
\eeq
\beq
<s|V_c|s>=\frac{2U}{N}
\eeq
\emath
Again the average of Eq. (25a) and Eqs. (25b) or (25c) yields Eq. (19) or (18) for $\bk'=-\bk$.

\section{relevance to superconductivity}

Our findings in  the previous sections suggest that the new basis is relevant to superconductivity. The fact that the BCS wavefunction Eq. (4)  singles out pairs $(\bold{k} \uparrow,-\bold{k}\downarrow)$
could naturally be explained by the fact that the direct Coulomb repulsion between an $\uparrow$-spin electron in state $\tilde{\varphi}_\bold{k}$ and a $\downarrow$-spin electron in
$\tilde{\varphi}_{-\bold{k}}$ is particularly small.
 
Consider then as an alternative to the  BCS wavefunction a wavefunction of the form Eq. (4) but with the new states instead of the Bloch states: \beq
|\Psi>_{new}=\prod_{\bold{k}} (u_{\bold{k}}+v_{\bold{k}} \tilde{c}^\dagger_{\bold{k}\uparrow} \tilde{c}^\dagger_{\bold{-k}\downarrow}) |0>
\eeq
Deep in the Fermi sea,  $u_\bold{k}=u_{-\bold{k}}=0$ and  $v_\bold{k}=v_{-\bold{k}}=1$. Since
\beq
(\tilde{c}^\dagger_{\bold{k}\uparrow}\tilde{c}^\dagger_{-\bold{k}\downarrow})
(\tilde{c}^\dagger_{-\bold{k}\uparrow}\tilde{c}^\dagger_{\bold{k}\downarrow})
=(c^\dagger_{\bold{k}\uparrow}c^\dagger_{-\bold{k}\downarrow})
(c^\dagger_{-\bold{k}\uparrow}c^\dagger_{\bold{k}\downarrow})
\eeq
deep in the Fermi sea the new wavefunction is the same as the BCS wavefunction. 
This also means that in the normal state ($u_\bold{k} v_\bold{k}=0$ for all $\bold{k}$)  the wavefunctions Eq. (4) and Eq. (26) are identical. However, in the region where $u_\bold{k} v_\bold{k}\neq 0$ there is a difference.
We have for $(\bold{k}\uparrow,-\bold{k}\downarrow)$ and $(-\bold{k}\uparrow,\bold{k}\downarrow)$ pairs in the BCS wavefunction
\beqn
|\Psi>_{BCS}^{\bold{k},-\bold{k}}&=&(u_\bold{k}+v_\bold{k}c^\dagger_{\bold{k}\uparrow}c^\dagger_{-\bold{k}\downarrow})(u_{-\bold{k}}+v_{-\bold{k}}c^\dagger_{-\bold{k}\uparrow}c^\dagger_{\bold{k}\downarrow})
\nonumber \\
&=& 
u_\bold{k}^2 + v_\bold{k}^2
(\tilde{c}^\dagger_{\bold{k}\uparrow}\tilde{c}^\dagger_{-\bold{k}\downarrow})
(\tilde{c}^\dagger_{-\bold{k}\uparrow}\tilde{c}^\dagger_{\bold{k}\downarrow}) \nonumber \\
&-&i u_\bold{k} v_\bold{k}((\tilde{c}^\dagger_{\bold{k}\uparrow}\tilde{c}^\dagger_{\bold{k}\downarrow})+(\tilde{c}^\dagger_{-\bold{k}\uparrow}\tilde{c}^\dagger_{-\bold{k}\downarrow}))
\eeqn
(assuming as usual $u_\bold{k}=u_\bold{-k}$, $v_\bold{k}=v_\bold{-k}$). Instead, in the new wavefunction Eq. (26) 
\beqn
|\Psi>_{new}^{\bold{k},-\bold{k}}&=&
(u_\bold{k}+v_\bold{k}\tilde{c}^\dagger_{\bold{k}\uparrow}\tilde{c}^\dagger_{-\bold{k}\downarrow})(u_{-\bold{k}}+v_{-\bold{k}}\tilde{c}^\dagger_{-\bold{k}\uparrow}\tilde{c}^\dagger_{\bold{k}\downarrow})
\nonumber \\
&=& 
u_\bold{k}^2 + v_\bold{k}^2
(\tilde{c}^\dagger_{\bold{k}\uparrow}\tilde{c}^\dagger_{-\bold{k}\downarrow})
(\tilde{c}^\dagger_{-\bold{k}\uparrow}\tilde{c}^\dagger_{\bold{k}\downarrow}) \nonumber \\
&+&u_\bold{k} v_\bold{k}((\tilde{c}^\dagger_{\bold{k}\uparrow}\tilde{c}^\dagger_{\bold{-k}\downarrow})+(\tilde{c}^\dagger_{-\bold{k}\uparrow}\tilde{c}^\dagger_{\bold{k}\downarrow}))
\eeqn
Eqs. (28)  and  (29) differ in the last line. The BCS wavefunction has an amplitude for having double occupancy of the state $\tilde{\varphi}_\bold{k}$ 
with the state $\tilde{\varphi}_\bold{-k}$ being empty, and vice versa, and no amplitude for the states $\tilde{\varphi}_\bold{k}$ and
$\tilde{\varphi}_\bold{-k}$ being both singly occupied . Instead, the new state has amplitude for the states $\tilde{\varphi}_\bold{k}$ and $\tilde{\varphi}_\bold{-k}$  being  both singly-occupied and no amplitude for the state $\tilde{\varphi}_\bold{k}$ being doubly occupied with the state $\tilde{\varphi}_\bold{-k}$
empty, nor vice versa. Single occupancy of both $\tilde{\varphi}_\bold{k}$ and $\tilde{\varphi}_\bold{-k}$ minimizes the direct Coulomb repulsion between the electrons while double occupancy of either state maximizes the direct Coulomb repulsion.
Consequently, we conclude that the new wavefunction Eq. (26) will be strongly favored energetically over the BCS wavefunction Eq. (4) as far as the
Coulomb interaction between electrons is concerned.

In summary, in the normal state where states $\bk$ and $-\bk$ are both either occupied or empty, the new basis is completely equivalent to the conventional one. However in a state of the BCS form that allows for partial
occupation of pair states the new basis appears to be favorable. 

Furthermore, Eqs. (29), (16b) and (14b) indicate that the advantage of the new basis over the conventional one will be greatest if $u_\bk  v_\bk \ne 0$ in a region where $k$ is small. 
Instead, for $\bk$ near the edge of the Brillouin zone there is no advantage to the new basis according to Eq. (19). This suggests that in a superconducting state described by the
wavefunction Eq. (26) the region where $u_\bk v_\bk\neq 0$ should occur near $k=0$, and the states with $\bk$ near the Brillouin zone edge should be full
($v_\bk=1, u_\bk=0$) (to conserve the number of particles). This is precisely the scenario predicted in Ref.\cite{holeelectron3}.

\section{4-electron state}
We consider in the following the plane wave case only. Let us consider further  the two candidate wavefunctions Eqs. (4) and (26)  for the case of only two pairs,   in states 
$(\bold{k}\uparrow,-\bold{k}\downarrow)$ and $(-\bold{k}\uparrow,\bold{k}\downarrow)$. The Coulomb energy in the BCS case is
\bmath
\beq
<V_c>_{BCS}=\frac{4V(0)}{\Omega} |v_\bk|^4 +\frac{2 (V(0)+V(2k))}{\Omega} |u_\bk|^2 |v_\bk|^2  
\eeq
and with the new wavefunction Eq. (26) it is
\beq
<V_c>_{new}=\frac{4V(0)}{\Omega} |v_\bk|^4 +\frac{2V(0) }{\Omega} |u_\bk|^2 |v_\bk|^2 
\eeq
\emath
Since the Coulomb repulsion $V(2k)$ is positive for all $k$, the new state has clearly lower Coulomb energy than the BCS state.

However in adding the second electron pair we have lost the lowering of  energy proportional to $V(2k)$ that occurred for the single pair (Eq. (16b)) due to interference effects.
How can it be restored? In deriving Eq. (30), we assumed $u_\bk=u_{-\bk}$, $v_\bk=v_{-\bk}$. Not making that assumption we obtain instead
\bmath
\beqn
&<&V_c>_{BCS}=\frac{V(0)}{\Omega}   (|v_\bk|^2+|v_{-\bk}|^2)^2 \nonumber \\
&+&\frac{V(0)}{\Omega} (|u_\bk|^2 |v_\bk|^2 +|u_{-\bk}|^2 |v_{-\bk}|^2) \nonumber \\
&+& \frac{V(2k)}{\Omega} (u_\bk^*v_\bk u_{-\bk}v_{-\bk}^*+u_{-\bk}^*v_{-\bk}u_\bk v_\bk^*)
\eeqn
and with the new wavefunction Eq. (26)
\beqn
&<&V_c>_{new}=\frac{V(0)}{\Omega}   (|v_\bk|^2+|v_{-\bk}|^2)^2 \nonumber \\
&+&\frac{V(0)}{\Omega} (|u_\bk|^2 |v_\bk|^2 +|u_{-\bk}|^2 |v_{-\bk}|^2) \nonumber \\
&-&\frac{V(2k)}{2\Omega} |u_\bk v_{-\bk}-u_{-\bk}v_\bk|^2
\eeqn 
\emath
The last term in Eq. (31b) lowers the energy if
\beq
\frac{v_\bk}{u_\bk}\neq
\frac{v_{-\bk}}{u_{-\bk}}
\eeq
that is, if the occupation of $(\tilde{\bk}\uparrow,-\tilde{\bk}\downarrow)$ is different from that of $(-\tilde{\bk}\uparrow,\tilde{\bk}\downarrow)$ .
Thus, it favors a spontaneous breaking of parity and the existence of a spin current\cite{sc}. 

In earlier work we have presented several  arguments in favor of the suggestion that superconductors possess a spontaneous
spin current in their ground state\cite{sm,atom,lorentz,sc}.
The direction of the spin current is determined by the spin-orbit interaction in the presence of an outward-pointing electric field 
predicted by the theory of hole superconductivity\cite{chargeexp} together with the sample
geometry.
Namely, near a surface with outward normal $\bold{\hat{n}}$     the member of the Cooper pair of spin $\bold{\sigma}$ has larger velocity in
direction $\bold{\hat{n}}\times \vec{\sigma}$\cite{sc}. Also, C.R. Hu has argued quite generally that the structure of the BCS state is compatible with the existence of a spin current\cite{hu}.

\section{Coulomb interaction in the pair wavefunctions}
The average of the Coulomb interaction in the BCS wavefunction Eq. (4)  is given by
\beq
<V_c>_{BCS}=\frac{1}{\Omega}\sum_{\bk \bk'}u_\bk^*v_\bk u_{\bk'}v^*_{\bk '}V(\bk-\bk')+\frac{V(0)}{\Omega}(\sum_\bk|v_k|^2)^2
\eeq
and in the new wavefunction Eq. (26) by
\beqn
<V_c>_{new}&=&\frac{1}{2\Omega}\sum_{\bk \bk'}u_\bk^*v_\bk u_{\bk'}v^*_{\bk '}[V(\bk'-\bk)-V(\bk'+\bk)] \nonumber \\
&+& \frac{V(0)}{\Omega}(\sum_\bk|v_k|^2)^2 \nonumber \\
&+& \frac{V(0)}{2\Omega}\sum_\bk (u_\bk^*v_\bk u_{-\bk}v^*_{-\bk }+|u_\bk|^2|v_\bk|^2))\nonumber \\
&+& \frac{1}{2\Omega} \sum_\bk V(2k)|v_\bk|^2[|v_{-\bk}|^2-|v_\bk|^2)
\eeqn
For a clearer comparison we separate self-energy terms in the double sums and obtain for the BCS case
\beqn
&<&V_c>_{BCS}=\frac{1}{\Omega}\sum_{\bk \neq \bk'}u_\bk^*v_\bk u_{\bk'}v^*_{\bk '}V(\bk-\bk')  \nonumber \\
&+&\frac{V(0)}{\Omega}(\sum_\bk|v_k|^2)^2 +\frac{V(0)}{\Omega} \sum_k |u_\bk|^2|v_\bk|^2
\eeqn
and for the new wavefunction
\beqn
&<&V_c>_{new}=\nonumber \\
& &\frac{1}{2\Omega}\sum_{\substack{\bk\neq \bk'\\ \bk\neq -\bk'}}   u_\bk^*v_\bk u_{\bk'}v^*_{\bk '}[V(\bk'-\bk)-V(\bk'+\bk)] \nonumber \\
&+&\frac{V(0)}{\Omega}(\sum_\bk|v_k|^2)^2 +\frac{V(0)}{\Omega} \sum_k |u_\bk|^2|v_\bk|^2 \nonumber \\
&-&\frac{1}{4\Omega} \sum_\bk V(2k) |u_\bk v_{-\bk}-u_{-\bk}v_\bk|^2
\eeqn

The BCS expression Eq. (35) has the same form whether or not parity is broken. Instead, in the expression Eq. (36) both the first and last term vanish identically
if parity is unbroken, and the average of the Coulomb interaction (as well as of any other interaction) gives a self-energy which is identical to that of the BCS case
(2nd lines in Eqs. (35) and (36)).

Thus we conclude that there is a stark difference between the BCS wavefunction and the new wavefunction. The BCS wavefunction will give rise to superconductivity
if the first term in Eq. (35) is negative on average\cite{bcs}. This will be the case if the interaction $V(\bk-\bk')$ in addition to the repulsive Coulomb interaction contains
negative (attractive) terms arising e.g. from the electron-phonon interaction, $and$ if the attractive terms are sufficiently large that they can overwhelm the repulsive
Coulomb terms. Of course the frequency dependence of the electron-phonon interaction provides additional help by reducing the effective strength of the 
Coulomb interaction\cite{andersonmorel}.

Instead, with the new wavefunction an attractive interaction (whether instantaneous or retarded)  will not help if parity is unbroken, since it will cancel identically in the first term in Eq. (36).
The direct Coulomb repulsion is however also eliminated   in the first term in Eq. (36) in contrast to the first term in Eq. (35) if parity is unbroken. Furthermore,
 if parity is broken, the last term in Eq. (36) provides a lowering of energy in the superconducting state {\it for a repulsive interaction},
 and particularly if the region where $u_\bk v_\bk \neq 0$ occurs near $k=0$.

\section{discussion}

It is natural to try to describe the superconducting state using electronic states that do not optimize the electron-ion interaction and many such attempts have been tried
in the past. For example, Froehlich\cite{froelich} proposed a state where  a shell of occupied states at the top of the normal Fermi distribution is moved outward, leaving behind an empty
shell.   This has a certain resemblance to the scenario proposed in Ref. \cite{holeelectron3}.   Even in conventional BCS theory, the normal state Fermi surface is `blurred' and Bloch states that are empty in the normal state
(i.e. unfavorable for the electron-ion interaction) become partially occupied in the BCS state. However we are not aware of earlier attempts to describe the 
superconducting state using single particle states that violate Bloch's theorem as we propose in this paper, except for
the early work of Kronig\cite{kronig} and
 Heisenberg\cite{heisenberg} that proposed theories of superconductivity based on an electronic `lattice' 
 that had no relation to the underlying ionic lattice.
In addition, Ref.\cite{tanaka}  used sine functions to take into account the effect of sufaces and impurities.

As discussed here, the basis introduced in this paper is consistent with the proposal of ref. \cite{holeelectron3} that in the superconducting state the states near the top of the
band / edge of the Brillouin zone become full and those near the bottom of the band / center of the zone, that resemble free-electron-like states, become empty or
partially occupied (Fig. 6 of ref. \cite{holeelectron3}). This is because the new basis gives rise to energy lowering compared with the conventional basis for small $\bk$ (near the bottom of the band),
while it has no advantage over the conventional basis for $\bk$ near the edge of the Brillouin zone (states near the top of the band).    This in turn is consistent with the observation that properties of the superconducting state are `universal', because they arise from long wavelength states near the bottom of the band  that are
insensitive to details of the electron-ion potential. In order for a system to be able to push the  electronic occupation to the top of the band  (Fig. 6 of ref. \cite{holeelectron3}) without undue energy cost requires the band
to be almost full in the first place (hole conduction in the normal state), and is favored by a low curvature of the band near the top (large effective mass), a narrow band and a small
ionic charge $Z$, which are all requirements that we discussed earlier in connection with the theory of hole superconductivity but based on different arguments.

We also found here that with the new basis the energy is lowered if parity is broken and a spin current exists, consistent with our earlier predictions\cite{sc,sm}.
In a small cluster, a rigid electronic state where parity is broken should lead to an electric dipole moment as observed in the remarkable experiments of de Heer and coworkers\cite{deheer}.

The new basis proposed here may be of interest also for other unconventional theories of superconductivity besides the theory of hole superconductivity.

 \end{document}